\documentclass[prl,a4paper,twocolumn,showpacs]{revtex4}

\usepackage{dcolumn}
\usepackage{color}
\usepackage{bm}
\usepackage{graphicx}
\usepackage{epsfig}
\usepackage{amsmath}
\usepackage{amsfonts}
\usepackage{amssymb}
\newcommand{\beq}{\begin{equation}}
\newcommand{\eeq}{\end{equation}}
\newcommand{\eqna}{\begin{eqnarray}}
\newcommand{\eqne}{\end{eqnarray}}
\newcommand{\mb}{\mathbf}
\newcommand{\vecx}{\mathbf{x}}

\begin{document}

\title{Path Integrals for Photonic Crystals.}

\author{Yair Dimant and Shimon  Levit\footnote{The Harry and Kathleen Kweller Professor of Condensed Matter Physics.} ,}
\affiliation{Department of Condensed Matter Physics, The Weizmann Institute of Science, Rehovot 76100, Israel}
\date{\today}

\begin{abstract}
We develop a path integrals approach for  analyzing stationary light propagation
appropriate for photonic crystals. The hermitian form of the stationary Maxwell equations is transformed into
a quantum mechanical problem of a spin 1 particle with spin-orbit coupling and position dependent mass.
After appropriate ordering several path integral representations of a solution are constructed. One leaves the propagation of polarization degrees of freedom in an operator form integrated over paths in coordinate space. The use of spin 1 coherent states allows to represent this part as a path integral over such states.  Finally a path integral in transversal momentum space with explicit transversality enforced at every time slice is also given.  As an example the geometrical optics  limit is discussed
and the ray equation is recovered together with the Rytov rotation of the polarization vector.
\end{abstract}
\maketitle

{\em Introduction.} Description of stationary light fields in photonic crystals can be formulated,
cf. \cite{photonic-crystals1,photonic-crystals2},   as solutions of an eigenvalue equation,
\beq \label{mastereq}
 \nabla\times\frac{1}{\varepsilon(\mathbf{x})}\nabla\times\mathbf{H}=
 \left(\frac{\omega}{c}\right)^{2}\mathbf{H}
\eeq
 where $\mathbf{H}$ is the magnetic field and $\epsilon(\vecx)$ is the space dependent
 dielectric constant.   This equation is sometimes referred to in the literature as the master
equation. The operator $\hat{\Theta}=\nabla\times\frac{1}{\varepsilon(\mathbf{x})}\nabla\times$
is hermitian and positive definite. This formulation allows direct applications to photonic crystals
of  many powerful techniques developed in  quantum mechanics.

Our goal is to follow on this development and formulate the path integral representation of solutions
 of this equation. Within the scalar approximation
\beq
-\frac{1}{\varepsilon(\mathbf{x})}\nabla^{2}u(\mathbf{x})=\left(\frac{\omega}{c}\right)^{2}\cdot u(\mathbf{x})
\eeq
 works in this spirit have already been presented in the past \cite{scalar1,scalar2,scalar3,scalar4,schul}, but here we deal with the full vector version.  To the best of our knowledge this have
not yet been done.

Let us consider an auxiliary time dependent Schrodinger like equation
\beq \label{auxScheq}
i\partial_{\tau}\mathbf{C}\left(\mathbf{x},\tau\right)=
\hat{\Theta}\mathbf{C}\left(\mathbf{x},\tau\right)
\eeq
with fictitious time parameter $\tau$. This equation has the formal solution
 \beq \label{formalsolution}
\mathbf{C}(\mathbf{x},\tau)=
\exp\left[-i\tau\hat{\Theta}\right]\mathbf{C}(\mathbf{x},0)
\eeq
from which we can recover a solution of the original master equation by using
\beq \label{intrelation}
\mathbf{H}(\vecx)=\lim_{\eta\rightarrow 0} \int_{-\infty}^{\infty} d\tau\left[e^{i\left(\frac{\omega}{c}\right)^{2}\tau-\eta|\tau|}\mathbf{C}
(\mathbf{x},\tau)\right]
\eeq
 Note that for any $\omega\ne 0$ the solutions $\mb{H}(\vecx)$ of Eq. (\ref{mastereq})
are transversal, $\nabla\cdot\mb{H}=0$. The solutions $\mathbf{C}\left(\mathbf{x},\tau\right)$ of
 (\ref{auxScheq}) are not. However since $\partial_\tau [\nabla\cdot \mathbf{C}(\vecx,\tau)]=0$ we have
  that the above transversality condition is satisfied also
  by $\mb{H}(\vecx)$ given by (\ref{intrelation}).

{\em The Dirac notations.} The Hilbert space of all complex vector
 functions $\mb{H(\vecx)}$  is equivalent to the Hilbert space of a spin 1 particle in
quantum mechanics. The spin operators are hidden in the vector product signs which we make explicit by using the antisymmetric  tensor (and the summation over repeated indices convention)
\eqna
\left[\hat{\Theta}\mathbf{H}\right]_{i}&=&
\left[\nabla\times\left(\frac{1}{\varepsilon(\vecx)}
\nabla\times\mathbf{H}\right)\right]_{i}= \nonumber \\
&=&\epsilon_{ijk}\epsilon_{klm}\;\partial_{j}\,\frac{1}{\varepsilon(\vecx)}\,\partial_{l}\,H_{m}= \\
&=&(-i \epsilon_{ikj})(-i \epsilon_{kml})
\left(-i\partial_{j}\right)\frac{1}{\varepsilon(\vecx)}\,\left(-i\partial_{l}\right) H_{m}, \nonumber
\eqne
 The operator $\hat{\Theta}$ is therefore
\beq \label{Theta}
\hat{\Theta}=(\hat{\mb{p}}\cdot\hat{\mb{S}}) v
(\vecx)(\hat{\mb{p}}\cdot\hat{\mb{S}})
\eeq
with $\hat{\mb{p}}=-i\nabla$ ,
$$
v(\vecx)=1/\varepsilon(\vecx)
$$
 and $\hat{\mb{S}}$
-- the spin 1 matrix vector with components $[\hat{S}_j]_{ik}=-i \epsilon_{ikj}$.

 It is convenient to introduce the Dirac notations
\eqna
\hat{\Theta}\left|H\right\rangle &=&\left(\frac{\omega}{c}\right)^{2}\left|H\right\rangle  \\
H_{i}\left(\mathbf{x}\right)&=&\left\langle \mathbf{x},i|H\right\rangle , \quad i=1,2,3 \nonumber \\
\langle \mathbf{x},i|\hat{\Theta}|H \rangle &=&\left[\nabla\times\frac{1}{\varepsilon(\mathbf{x})}\nabla\times\mathbf{H}
\left(\mathbf{x}\right)\right]_{i} \nonumber
\eqne
In this notation equation (\ref{formalsolution}) is
\beq
\langle \vecx, i|\mb{C}(\tau)\rangle =\sum_{j=1}^3\int d\vecx' \langle \vecx,i| e^{-i\tau\hat{\Theta}}|\vecx',j\rangle\langle \vecx',j|\mathbf{C}(0)\rangle
\eeq

{\em Ordering the "Hamiltonian".} We intend to write the path integral expression for the
  propagator $\langle \vecx,i| e^{-i\tau\hat{\Theta}}|\vecx',j\rangle$.
 Following the usual procedure \cite{ordering} we rewrite $\hat{\Theta}$ in the ordered form placing the
 $\hat{\mb{p}}$ operators to the right of $\mb{x}$'s,
 \beq \label{orderedtheta}
 \hat{\Theta}=-i(\nabla v\cdot\hat{\mb{S}})(\hat{\mb{p}}\cdot\hat{\mb{S}}) +v(\vecx)\hat{\mb{p}}^{2}\hat{\mathcal{P}}_{T}
 \eeq
 with  $\hat{\mathcal{P}}_{T}$ -- the transversal projection operator
\beq  \label{projoper}
[\hat{\mathcal{P}}_{T}]_{ij}=
\left[\frac{(\hat{\mb{p}}\cdot\hat{\mb{S}})^{2}}{\hat{\mb{p}}^{2}}\right]_{ij}=
\delta_{ij}-\frac{\nabla_i\nabla_j}{\nabla^2}
\eeq
 where we used  $(\hat{S}_i\hat{S}_j)_{mn}=\epsilon_{kmi}\epsilon_{knj}=\delta_{ij}\delta_{mn}-\delta_{mj}\delta_{ni}$
We are interested only in how the transverse functions $\mb{C}(\vecx,\tau)$ propagate for
 which $\hat{\mathcal{P}}_{T}|\mb{C}\rangle=|\mb{C}\rangle$. Accordingly we can drop $\hat{\mathcal{P}}_{T}$
 in Eq. (\ref{orderedtheta}) and define  a reduced operator
\beq \label{reducedTheta}
\hat{\Theta}_R=v(\vecx)\hat{\mb{p}}^{2}-i(\nabla v\cdot\hat{\mb{S}})(\hat{\mb{p}}\cdot\hat{\mb{S}})
\eeq
Note that the operators (\ref{Theta}) and (\ref{reducedTheta}) are equivalent when acting on transversal functions. In general one can show  that
$\hat{\Theta}=\hat{\Theta}_{R}\mathcal{P}_{T}$. Note that $\hat{\Theta}_{R}$ is hermitian in the transverse subspace.

{\em Path Integral - The Operator Version.} We are now in a position to use the standard time slicing process to construct the path integral.  In this we first choose to keep the vector part of the evolution at the operator level inserting complete (coordinate and momentum) states only for the spatial part. As a result we obtain a functional integral over the positions and momenta of spatial paths with the integrand containing the evolution operator of the vectorial part of the field for each path
\begin{widetext}
\eqna  \label{operversion}
\langle \vecx| e^{-i\tau\hat{\Theta}}|\vecx'\rangle&=&
\lim_{\Delta\tau\rightarrow 0}\int\frac{d\mathbf{x}_{1}d\mathbf{p}_{1}\ldots d\mathbf{p}_{N}}{\left(2\pi\right)^{3N}}\prod_{j=1}^{N}e^{i\mathbf{p}_{j}\cdot (\vecx_{j}-\vecx_{j-1})}\left[1-i\Delta\tau\, v_{j}\,\mathbf{p}_{j}^{2}\right]\left[1-\Delta\tau(\nabla v_{j}\cdot\hat{\mb{S}})(\mathbf{p}_{j}\cdot\hat{\mb{S}})\right]
\nonumber \\ &=&\int
D[\vecx(\tau),\mb{p}(\tau)]\exp\left[i\int_{0}^{\tau}\left(\mb{p}\cdot\partial_{\tau}
\vecx-v\mb{p}^{2}\right)d\tau\right]T\left\{ \exp{[-\int_{0}^{\tau}(\nabla v \cdot \hat{\mb{S}})(\mb{p}\cdot {\hat{\mb S}})d\tau }]\right\}
\eqne
\end{widetext}
where $T$ stands for time ordering, $\Delta \tau=\tau/N$, $\vecx(0)=\vecx'$, $\vecx(\tau)=\vecx$. This functional integral has position dependent mass $m(\vecx)=1/2v(\vecx)=\epsilon(\vecx)/2$ and spin orbit  coupling terms which enter the time ordered exponential.

{\em The Spin Coherent States Version.} The operator part of the above path integral, i.e. the time ordered exponential can be further developed using the spin coherent states. These states are generated, cf. \cite{auerbach,k1,k2,simons}, by rotations of one of the eigenstates of $\hat{S}_z$, i.e.  $|1,\mu\rangle$, $\mu=1,0,-1$. In the cartesian basis used above these states are
\beq
\left|1,1\right\rangle =\frac{1}{\sqrt{2}}\left(\begin{array}{c}-1\\-i\\0\end{array}\right) \left|1,0\right\rangle =\left(\begin{array}{c}0\\0\\1\end{array}\right)\left|1,-1\right\rangle =\frac{1}{\sqrt{2}}\left(\begin{array}{c}+1\\-i\\0\end{array}\right) \nonumber
\eeq
The most common spin coherent states are given by
\beq
\left|\Omega\right\rangle =\left|\theta,\phi\right\rangle =e^{-i\phi \hat{S}_z}e^{-i\theta \hat{S}_y}\left|1,1\right\rangle
\eeq
They are eigenstates of the spin component in the appropriate direction,
$\;\mb{\Omega}\cdot\hat{\mb{S}}|\Omega\rangle=|\Omega\rangle $,
and form an over complete set
which has a useful "resolution of unity" property
\beq \label{resolunity}
\hat{I}=\frac{3}{4\pi}\int d\Omega\left|\Omega\left\rangle\right\langle \Omega\right|\;,\: d\Omega=\sin\theta d\theta d\phi .
\eeq
Going again  through the time slicing
process and inserting (\ref{resolunity}) between the slices one finds  the propagator between two spin coherent states
\begin{widetext}
\beq\label{cohstatesprop}
 \langle \mb{x}_{f},\Omega_{f}|e^{-i\tau\hat{\Theta}_{R}}|\mb{x}_{i},
 \Omega_{i}\rangle =
\int D\left[\mb{x},\mb{p}\right]\int D\left[\Omega\right] \exp\left[i\int_{0}^{\tau}\left(\mb{p}\cdot d_\tau\mb{x}+
\cos\theta\,d_\tau\phi-v\,\mb{p}^{2}+i
\langle \Omega|(\nabla  v\cdot\hat{\mathbf{S}})(\mb{p}\cdot\hat{\mb{S}})|\Omega\rangle\right) d\tau\right]
\eeq
\end{widetext}
All that is left is to evaluate the matrix element
\beq
\langle\Omega|(\nabla v\cdot \hat{\mathbf{S}})(\mathbf{p}\cdot\hat{\mb{S}})|\Omega\rangle =\sum_{i,j}(\nabla_{i}v)\, p_{j}\langle \Omega|
\hat{S}_{i}\hat{S}_{j}|\Omega\rangle \nonumber
\eeq
Using
$
\hat{S_{i}}\hat{S_{j}} =\frac{1}{2} ([\hat{S_{i}},\hat{S_{j}}] +
 \{ \hat{S_{i}},\hat{S_{j}} \} )\nonumber
$
together with the commutation relations of spin operators, the identity
\beq
\hat{S_{l}}\hat{S_{m}}\hat{S}_{n}+\hat{S_{n}}\hat{S_{m}}\hat{S_{l}}
=\delta_{l,m}\hat{S}_{n}+\delta_{n,m}\hat{S_{l}}
\eeq
valid for spin 1 matrices, \cite{spin1id} as well as
$$
\langle\Omega|\hat{S}_i|\Omega\rangle=\Omega_i,
$$
 cf. \cite{simons}, we obtain
\beq
\langle \Omega|\hat{S_{i}}\hat{S_{j}}|\Omega\rangle =\frac{i}{2}\epsilon_{ijk}\Omega_{k} +
\frac{1}{2}\left(\Omega_{i}\Omega_{j}+\delta_{ij}\right)
\eeq
and finally
\eqna
&&\langle \Omega|(\nabla v\cdot\hat{\mathbf{S}})(\mathbf{p}\cdot\hat{S})|\Omega\rangle =\frac{i}{2}\mb{\Omega}\cdot(\nabla v\times\mathbf{p}) + \nonumber \\
&&+\frac{1}{2}\left[(\nabla v\cdot\mathbf{\Omega})(\mathbf{p}\cdot\mb{\Omega})+\nabla v\cdot \mathbf{p}\right].
\eqne

{\em Explicit Transversal Projection.}   The above path integral expressions for the exact propagator will
 propagate only transverse vector functions despite unrestricted integrations at every time slice. It may be desirable however, especially in making approximations, to have path integral expressions with explicit transversal
 projection enforced at every time step. This can be achieved by inserting the projection operator (\ref{projoper}) at every time slice. Moreover one must start with a transverse state, the simplest of which is
  $\mbox{\boldmath $\lambda$}\exp(i\mb{p}\cdot\vecx)$ with $\mbox{\boldmath $\lambda$}\cdot\mb{p}=0$.
It is natural then to take also the final state to be a transversal plane wave.
Using the spin coherent states this means working with $\langle \mb{p}_{f},\Omega_{f}|e^{-i\tau\hat{\Theta}_{R}}|\mb{p}_{i},
 \Omega_{i}\rangle $ and imposing the transversality conditions
\beq
 \Omega_{i,f}=\pm\frac{\mathbf{p}_{i,f}}{\left|\mathbf{p}_{i,f}\right|}
\eeq
under which  $\mathcal{P}_{T}|\mb{p}_{i,f},
 \Omega_{i,f}\rangle=|\mb{p}_{i,f}, \Omega_{i,f}\rangle$.
 Once one uses such transversal initial and final states  and due to the fact that the transversality is conserved by the infinitesimal transformations in between infinitesimal time slices, the projection operator insertion will not alter the propagation.
Expression (\ref{operversion}) will then become (in the limit $\Delta\tau\rightarrow 0$)
\eqna
&&\langle \mb{p}_{f}|e^{-i\tau\hat{\Theta}_{R}}|\mb{p}_{i}\rangle = \nonumber \\
&&=\int D\left[{\bf \mb{p},\mb{x}}\right]\exp[-i\int_{0}^{\tau}\left(\mb{x}\cdot d_\tau\mb{p}+v\mb{p}^{2}\right)d\tau] \times\nonumber \\
&& \times\prod_{\tau}\left[1-\Delta \tau\left(\nabla v_{\tau}\cdot\hat{\mb{S}}\right)\left(\mb{p}_{\tau}\cdot\hat{\mb{S}}\right)\right]
\frac{\left(\mb{p}_{\tau}\cdot\hat{\mb{S}}\right)^{2}}{\mb{p}_{\tau}^{2}} \nonumber
\eqne

{\em The Saddle Point Approximation - The Geometrical Optics Limit.} Geometrical optics is a short wavelength expansion. The
vacuum wavelength of light must be short with respect to the scale $v/|\nabla v|$ over
which the dielectric function is changing. Accordingly we treat the spin dependent part of Eqs. (\ref{operversion}) and (\ref{cohstatesprop}) as slowly varying and evaluate it on the saddle
point of the functional $S_0 =
\int_{0}^{\tau}(\mb{p}\cdot d_\tau\vecx-v\mb{p}^{2})d\tau $.
The corresponding Euler-Lagrange  equations are
\beq
d_{\tau}\vecx=2 v\mathbf{p} \;\;\;\; ; \;\;\;\;
d_{\tau}\mb{p}=-\mb{p}^{2}\mb{\nabla}v\;.
\eeq
The conserved "energy" is  $ E=v\,\mb{p}^{2} $
which we will use in order to reparametrize
\beq
d_{\tau}=2v\sqrt{E}\, d_t
\eeq
and obtain the standard differential equation of the ray
\beq \label{rayeq1}
\frac{d^{2}\mathbf{x}}{dt^{2}}=\nabla\left(\frac{n^{2}}{2}\right)
\eeq
Once the geometrical ray (or rays) satisfying
 the appropriate boundary conditions (e.g. $\vecx(0)=\vecx', \vecx(\tau)=\vecx$) is found the propagator along each ray is given
\beq \label{raypropagator}
K=T\left\{ e^{-\int_{0}^{\tau}\left(\nabla v_{r}\cdot\hat{\mb{S}}\right)\left(\mb{p}_{r}\cdot \hat{\mb{S}}\right)d\tau}\right\}
e^{iS_0(\vecx,\vecx')} \int D\left[\mathbf{x},\mathbf{p}\right]e^{i\delta^2 S_0/2}
\eeq
where the subscript $r$ means evaluation at the ray values which also applies for  $S_0$ and $\delta^2S_0$. The Gaussian integral over $\exp{i\delta^2 S_0/2}$ is evaluated in a standard way \cite{schul},
giving the Van-Vleck determinant factor
$
\sqrt{(1/2\pi)^3 \det\{\partial^2_{\vecx\vecx'} S_0(\vecx,\vecx')\}}.
$
The evolution of the vector degrees of freedom  along a geometrical ray is given by the equation
\beq
\frac{d|\lambda\rangle}{d\tau} =-\left(\nabla v_{r}\cdot\hat{\mb{S}}\right)\left(\mb{p_{r}}\cdot\hat{\mb{S}}\right)\left|\lambda\right\rangle
\eeq
as follows from the time ordered operator in (\ref{raypropagator}). In the usual vector
representation this is just an evolution of a vector $\mbox{\boldmath $\lambda$}$
\beq \label{evolvec}
\frac{d\mbox{\boldmath $\lambda$}}{d\tau}=\nabla v_{r}\times\left(\mb{p}_r\times\mbox{\boldmath $\lambda$}\right)
\eeq
with time dependent $\nabla v_{r}\equiv\nabla v(\vecx_r(\tau))$ and $\mb{p}_r(\tau)$ determined by the ray
trajectory $\vecx_r(\tau)$, $\mb{p}_r(\tau)$. We will now show that the orientation  of $\mbox{\boldmath $\lambda$}$ relative to the ray is governed by the Rytov equation for the polarization while its magnitude  fits one of the components of the energy flow equation along the ray, cf.  \cite{Born,Landau}.

\begin{figure}[h]
\centering
\includegraphics[scale=0.3]{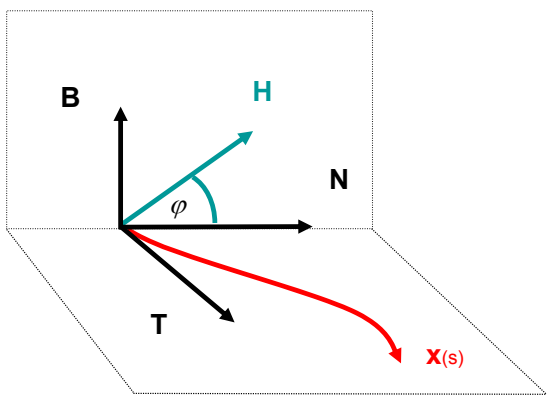}
\caption{Rytov rotation of the field relative to the ray.  $\mb{T}$, $\mb{N}$ and $\mb{B}$ are
respectively tangent, normal and binormal vectors. $\mb{H}$ is the magnetic field vector. $\mb{x}$ is the geometrical ray.}
\label{Rytov-figure}
\end{figure}

It is convenient to switch from "time" $\tau$ to the length parameter $s$  and write (\ref{evolvec}) as
\beq \label{equationinlength}
\frac{d\mbox{\boldmath $\lambda$}}{ds}=-\frac{\nabla n}{n}\times\left(\mathbf{T}(s)\times\mbox{\boldmath $\lambda$}\right)
\eeq
where $\mb{T}(s)=d\vecx/d s$ is the tangent to the curve. Considering also the normal and the binormal vectors
$$ \mathbf{N}(s)=\frac{d\mb{T}(s)}{ds}/\left|\frac{d\mb{T}(s)}{ds}\right|\;\;,\;\;
\mathbf{B}(s)=\mb{T}(s)\times\mb{N}(s)
$$
one can expand
\beq
\mbox{\boldmath $\lambda$}=\lambda_{T} \mathbf{T}+\lambda_{N} \mathbf{N}+\lambda_{B}\mathbf{B}.
\eeq
We transform the equation (\ref{rayeq1}) of the ray from $\tau$ to $s$
\beq
\frac{d}{ds}\left(n\frac{d\mathbf{x}}{ds}\right)=\nabla n
\eeq
and use the Frenet equations, describing the differential geometry of curves, \cite{Nov}
\beq
\frac{d}{ds}\left(\begin{array}{c}
\mathbf{T}\\
\mathbf{N}\\
\mathbf{B}\end{array}\right)=\left(\begin{array}{ccc}
0 & \kappa & 0\\
-\kappa & 0 & \tau\\
0 & -\tau & 0\end{array}\right)\left(\begin{array}{c}
\mathbf{T}\\
\mathbf{N}\\
\mathbf{B}\end{array}\right)
\eeq
where $\kappa\left(s\right)=|d\mb{T}/ds|$ is the curvature and $\tau\left(s\right)=\mb{B}\cdot d\mb{N}/ds$
is the torsion (not to be confused with the time parameter $\tau$).   We can rewrite Eq. (\ref{equationinlength}) as
\eqna
\frac{d}{ds}\left(\begin{array}{c}
\lambda_{T}\\
\lambda_{N}\\
\lambda_{B}\end{array}\right)=\left(\begin{array}{c}
0\\
-\kappa\lambda_{T}+\tau\lambda_{B}+\frac{d\ln{n}}{ds}\,\lambda_{N}\\
-\tau\lambda_{N}+\frac{d\ln{n}}{ds}\,\lambda_{B}\end{array}\right)
\eqne
The $\lambda_{T}\left(s\right)$ component is conserved along the
curve and we will take it to be zero (a transversal solution). We will rewrite the remaining two
equations as equations for $\varphi$ -- the angle between $\mbox{\boldmath $\lambda$}$ and  $\mb{N}$ and $\lambda^2=\lambda_{N}^2+
\lambda_{B}^2$-- the square length of $\mbox{\boldmath $\lambda$}$.  We obtain
\beq \label{Rytov}
\frac{d\varphi}{ds}=\tau\;\;\;,\;\;\;
\frac{d\lambda^{2}}{ds}=\frac{d\ln[n^2]}{ds}\lambda^{2}
\eeq
The first equation is the Rytov equation from geometrical optics governing the rotation
of polarization along an optical ray, cf., Fig. \ref{Rytov-figure}.  The solution of the second equation is simply
$ \lambda=n\left(s\right)\lambda_{0}$.

In the standard derivations  of the geometrical optics from the Maxwell equations, cf., \cite{Landau, Born},  one finds the following equation for the change with respect to the length parameter of the field amplitude square along a geometrical ray
\beq
\frac{d}{ds}\left|\mathbf{A}\right|^{2}=-
\left(\frac{\nabla^{2}S_{0}}{n}-\frac{d}{ds}\ln\left[n^{2}\right]\right)\left|\mathbf{A}\right|^{2}
\eeq
The two terms in the right hand side are clearly associated with the two contributions which we derived in the geometrical optics approximation to the amplitude of the propagator. The first term, which involves the second spatial derivatives of the action, is related to the Van-Vleck determinant contained in the last term in Eq. (\ref{raypropagator}).  The second term is identical  with what we obtained for $\lambda^{2}$ in Eq. (\ref{Rytov}).


\begin{thebibliography}{}
\bibitem{photonic-crystals1} J.D. Joannopoulos, S.Jhohnson, J. Winn, R. Meade, {\em Photonic Crystals - Modeling The Flow of light}, Princeton Univsersity Press, New Jersy (2008).
\bibitem{photonic-crystals2} J.D. Joannopoulos, P.R. Villeneuve, S. Fan, Nature {\bf 386} 143-149 (1997).
\bibitem{scalar1} M. Eve, Proc. Roy. Lond. A {\bf 347} 405-417 (1976).
\bibitem{scalar2} G. Samelsohn, R. Mazar, Phys. Rev. E {\bf 54} 5697 (1996).
\bibitem{scalar3} C. Gomez-Reino, J.Linares, J.Opt. Soc. Am. A {\bf 4} 1337 (1987).
\bibitem{scalar4} S.R. Vasta, J. Opt. Soc. Am. B {\bf 322} 2512 (2005).
\bibitem{schul} L.S. Schulman, {\em Techniques and Application of Path Integration},
Wiley, New York (1981).
\bibitem{ordering} T. Kashiwa, Y. Ohnuki, M. Suzuki, {\em Path Integral Methods}, Oxford University Press, (1997).
\bibitem{sc1} M. Pletyukhov, C. Amann, M. Metha and M. Brack, Phys.~Rev.~Lett.~{\bf 89} 116601-1 (2002).
\bibitem{sc2} M. Pletyukhov, O. Zaitsev, J.Phys. A: Math.Gen {\bf 36} 5181-5219 (2003).
\bibitem{sc3} C. Amann , M. Brack, J.Phys. A: Math.Gen {\bf 35} 6009-6032 (2002).
\bibitem{pwf1} I. Bialynicki-Birula, {\em Photon wave function}, Progress in Optics XXXVI , Amsterdam 245-294 (1996).
\bibitem{pwf2} I. Bialynicki-Birula, Acta  Phys. Polonica {\bf 86} 97-116 (1994).
\bibitem{auerbach} A. Auerbach, {\em Interacting Electrons and Quantum Magnetism}, Springer - Verlag, New York (1994).
\bibitem{k1} J.R. Klauder, B.S. Skagerstam,  {\em Coherent States. Applications in Physics and
 Mathmatical Physics}, World Scientific, Singapur (1985).
\bibitem{k2} J.R. Klauder, Phys. Rev. D {\bf  19} 2349 (1979).
\bibitem{simons} A. Altland, B. Simons, {\em Condensed Matter Field Theory}, Cambridge University Press (2006).
\bibitem{spin1id} I. Mendas, P. Milutinovic, J. Phys. A: Math. Gen. {\bf 23} 537-544, (1990).
\bibitem{Landau} L.D. Landau, E.M. Lifshitz {\em Electrodynamics of Continuous Media}, Pergamon press, Bristol (1960).
\bibitem{Born} M. Born, E. Wolf, {\em Principles of Optics}, Cambridge University Press (1959).
\bibitem{Keppeler} S. Keppeler, {\em Spining Particles - Semiclassics and Spectral Statistics}, Spinger, New York (2003).
\bibitem{GO} M. Kline, I.W. Kay, {\em Electromagnetic Theory and Geometrical Optics }, Wiely , New York (1965).
\bibitem{Nov} S. P. Novikov, A. T. Fomenko, {\em Basic Elements of Differential Geometry and Topology},
Kluwer, 1990.

\end{thebibliography}
\end{document}